\documentclass[a4paper, 10pt, conference]{IEEEtran}
\usepackage{cite}
\usepackage{amsmath,amssymb,amsfonts}
\usepackage{algorithmic}
\usepackage{graphicx}
\usepackage{textcomp}
\usepackage[export]{adjustbox}
\usepackage{xcolor}
\usepackage{multirow}
\usepackage[a4paper,left=0.7in,right=0.7in,top=0.8in,bottom=1.1in,]{geometry}

\def\BibTeX{{\rm B\kern-.05em{\sc i\kern-.025em b}\kern-.08em
    T\kern-.1667em\lower.7ex\hbox{E}\kern-.125emX}}
    
\def\vb(#1){
\mathbf{#1} }
    
\begin{document}

\title{Beam Profiling With Noise Reduction From Computer Vision and Principal Component Analysis for the MAGIS-100 Experiment
}

\author{\IEEEauthorblockN{Joseph Jachinowski}
\IEEEauthorblockA{\textit{Center for Fundamental Physics}\\
\textit{Department of Physics and Astronomy} \\
\textit{Northwestern University}\\
Evanston, IL 60208, USA \\
ORCID: 0000-0002-1053-5688}
\and
\IEEEauthorblockN{Natasha Sachdeva}
\IEEEauthorblockA{\textit{Center for Fundamental Physics}\\
\textit{Department of Physics and Astronomy} \\
\textit{Northwestern University}\\
Evanston, IL 60208, USA \\
ORCID: 0000-0001-9305-4208}
\and
\IEEEauthorblockN{Tim Kovachy}
\IEEEauthorblockA{\textit{Center for Fundamental Physics}\\
\textit{Department of Physics and Astronomy} \\
\textit{Northwestern University}\\
Evanston, IL 60208, USA \\
timothy.kovachy@northwestern.edu}
}

\maketitle
\thispagestyle{plain}
\pagestyle{plain}

\begin{abstract}

MAGIS-100 is a long-baseline atom interferometer that operates as a quantum sensor. It will search for dark matter, probe fundamental quantum science, and serve as a prototype gravitational wave detector in the 0.3 to 3~Hz frequency range. The experiment uses light-pulse atom interferometry where pulses of light create the atom optics equivalents of beamsplitters and mirrors. Laser beam aberrations are a key source of systematic error for MAGIS-100, and accurately characterizing the laser beam spatial profile is therefore essential. In this paper, we describe a new and efficient beam profiling technique. We use a low-cost CMOS camera affixed to a translating and rotating optomechanical mount to image the beam, then employ computer vision and principal component analysis to minimize background noise and produce accurate beam profiles for a laser incident on a variety of aberration-inducing optical elements.

\end{abstract}



\section{Introduction}

The Matter-wave Atomic Gradiometer Interferometric Sensor (MAGIS-100), currently under construction at Fermilab, is a 100-m tall atomic fountain that will be used to explore fundamental physics on a macroscopic scale. Functioning as a high-precision quantum sensor, it will search for ultralight dark matter and exotic forces, as well as test quantum mechanics in a new regime. MAGIS-100 will also serve as a testbed for new technologies required to eventually develop a km-scale atomic interferometer capable of detecting gravitational waves in the currently unexplored midband region, corresponding to 0.3 to 3 Hertz, bridging the gap between the peak sensitivities of LIGO and LISA \cite{Abe2021}.

In atom interferometry, atomic wavefunctions are made to coherently propagate over two paths and are then combined using a beamsplitter. The relative phase between the two paths is encoded in the probability of detecting the atoms at either output port. The phase shift contains information on the physical environment and relevant interactions to be studied. In light-pulse atom interferometry, the beamsplitters and mirrors along the interferometry paths are created by pulses of light. Aberrations in the atom optics laser beams often limit the sensitivity of the sensor \cite{tino2013atom}.

Laser beam aberrations impact atom phase through two mechanisms: (1) imprinting of the laser wavefront and (2) light shifts coupled to intensity perturbations. Diffraction and expansion of the atom cloud can cause the two interferometer arms to experience these phase shifts differently, ultimately resulting in dephasing of the atoms. 
These effects have been studied analytically \cite{Abe2021, Wicht2005, Gibble2006, Schkolnik2015} and would benefit from the addition of accurate simulations of beam propagation through the experimental apparatus as well as \textit{in situ} beam profiling. Moreover, intensity and phase perturbations are dynamically related and, thus, must be studied holistically \cite{Siegman_1986}. MAGIS-100 plans to use free-space propagation over four meters to clean the beam; therefore, only laser aberrations with a transverse spatial wavelength greater than approximately 8.8~mm will influence the interferometry sequence \cite{Abe2021}. 

Beam profiling is used to study laser beam aberrations and propagation. Many different techniques may be used to profile laser beams, including both non-electrical and electrical methods. Examples of non-electrical methods include burn paper, fluorescing plates, acrylic mode burns, and the human eye. Examples of electrical methods include knife-edge, slit, or pinhole imaging which measure the marginal power of the laser as a function of position, mechanically integrating over the entire beam profile. For such methods, diffraction around the knife, slit, or pinhole sets a lower limit on precision \cite{Dickey2000}.

Camera-based imaging systems are desirable due to their convenience, ease-of-use, and versatility. The primary image sensor options for a camera-based imaging system are a charge-coupled device (CCD) and a complementary metal oxide sensor (CMOS). While the significantly lower cost of cameras equipped with CMOS image sensors suggests a practical alternative to cameras equipped with CCD image sensors, CMOS technology also introduces additional sources of noise. CMOS image sensors are prone to both temporal and fixed pattern noise (FPN). FPN is non-Gaussian and requires advanced techniques to denoise \cite{Zhang2021}. On the other hand, because CMOS sensors can operate with a higher frame rate they are preferred over CCD sensors for video capture \cite{elgamal_2005_cmos, hain_2007_comparison}.

In addition to sensor noise, physical defects in optical elements such as surface defects on lenses, mirrors, and neutral density (ND) filters, as well as dust particles can result in diffraction fringes. In some cases, in order to properly characterize aberrations in the optical apparatus, such diffraction fringes are the primary focus of beam profiling efforts. However, extraneous aberrations, often a result of imperfect optical elements necessary for beam profiling, contribute to non-sensor noise. While non-sensor noise can be mitigated by using high-quality optical elements with precise specifications as well as operating in a dust-controlled environment, this is not always practical.

Sensor and non-sensor noise often appear stationary relative to the image boundary for a camera using a CMOS image sensor. If the camera is translated/rotated, then the desired image, in our case the beam profile, is also translated/rotated with respect to the image boundary. Stationary noise is invariant under such translation/rotations. Thus, if a series of images are captured during camera translations/rotations, then the appearance of the stationary noise relative to the beam profile will change. Using a video on the order of tens of seconds, a standard CMOS camera will be able to capture hundreds of images per video. Assuming that the beam profile is constant in time, post-processing of the image set can isolate the beam profile from the stationary noise. 

One such technique of post-processing is principal component analysis (PCA). PCA is a statistical method to reduce the dimensionality of a dataset while retaining the most information possible \cite{Shlens2014}. Applied to images, dimensionality reduction has been effectively implemented in denoising algorithms \cite{Muresan2003, Zhang2009, James2015, Zhang2021}. Moreover, PCA has been effectively used in a variety of applications across many fields in physics \cite{Jolliffe2016}. For example, PCA has been used extensively in digital holographic microscopy, an important tool in biomedical imaging \cite{Zuo2013, Zhang2019}, phase-shift interferometry \cite{Vargas2013, Escobar2020, Yatabe2016}, laser-induced breakdown spectroscopy \cite{Porizka2018}, imaging of ultra-cold atoms \cite{Segal2010, Cao2019, Xiong2020}, as well as measurements of atom number densities \cite{Ockeloen_2010}. PCA has also been used in atom interferometry to remove noise in images of atomic wave-packets \cite{Chiow2011, Dickerson2013}.

In this paper, we present a method for beam profiling which uses a camera-based imaging system equipped with a low-cost CMOS sensor to resolve accurate, low-noise beam profiles by translating and rotating the camera and post-processing with computer vision and principal component analysis (PCA). The technique removes noise from laser profile measurements to resolve features of aberrated beams, which enables the experimenter to identify a particular aberration of interest. We begin by briefly summarizing the method. Then, we provide a description of the experimental implementation and analysis and conclude with a discussion of our results.

\begin{figure}[thb]
    \includegraphics[width = 0.48\textwidth, left]{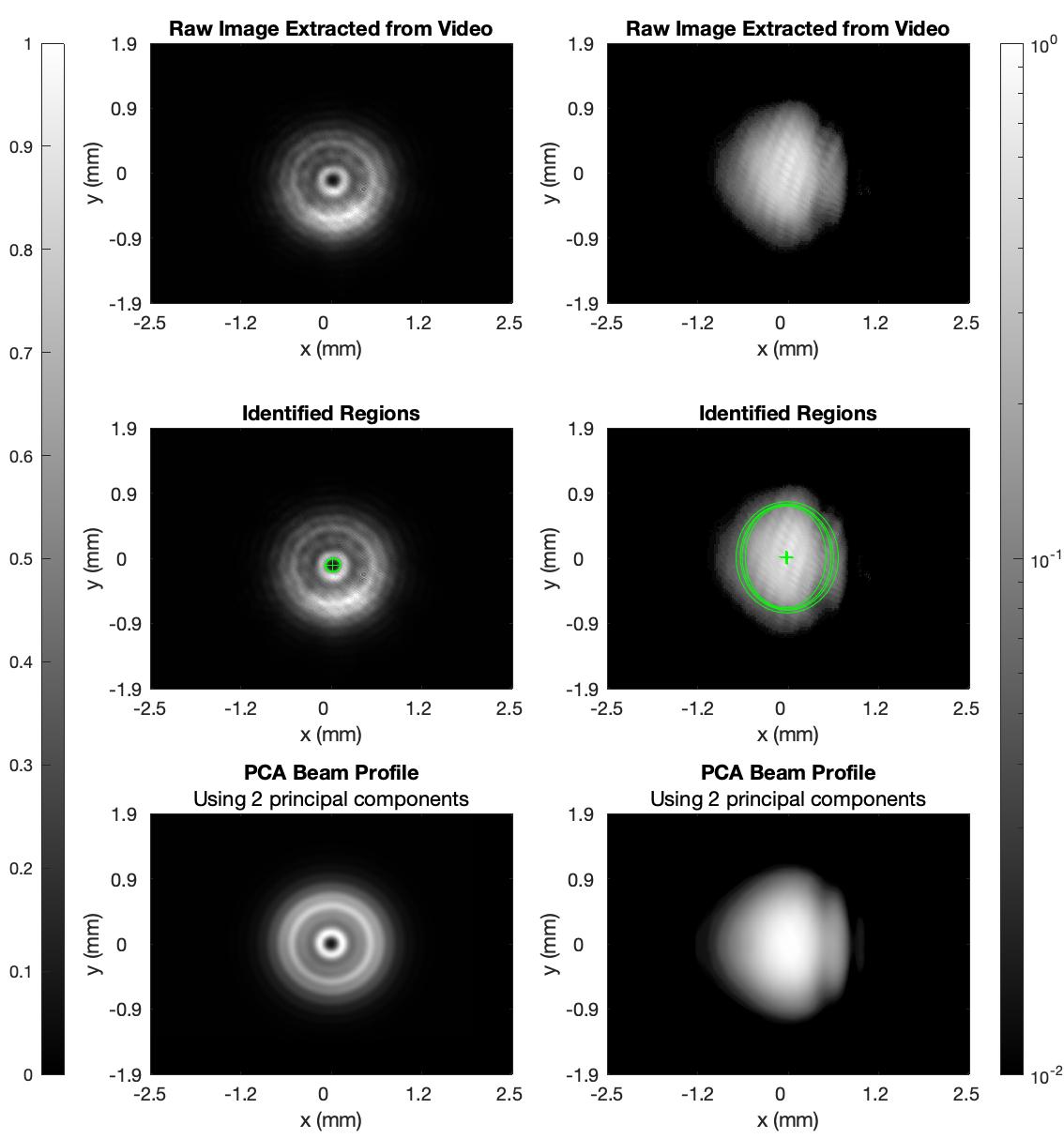}
    \caption{Region identification and PCA for two different beam profiles. The left beam profile corresponds to a laser incident on a 3.65~mm circular aperture propagating 177~cm. The right corresponds to a laser clipped by a hard-edge propagating 229~cm. For the beam profile on the right, a logarithmic color scale is used to show low-intensity details. The second row displays the identified "key regions" of the raw beam profiles circled in green. The third row displays the PCA beam profiles obtained using the first two PCs.}
    \label{example}
\end{figure}

We summarize the method as follows: (1) align camera detector with beam so that it is contained within the image boundaries, (2) begin video capture, (3) translate or rotate the camera while ensuring that the beam remains within image boundaries, (4) qualitatively identify key regions of beam profile from a sample frame as a precursor to (5) implementing a computer vision algorithm to identify key regions for the rest of the image dataset, (6) omit misidentified images and use identified regions to align the centers of the images, and (7) apply PCA and reconstruct beam profile using one to several principal components. Fig.~\ref{example} displays the implementation of this method for two different beam profiles. 

\section{Methods}\label{sec:methods}

\subsection{Experimental Setup}\label{sec:setup}

\begin{figure}[t!]
    \centering
    \includegraphics[width = 0.48\textwidth, center]{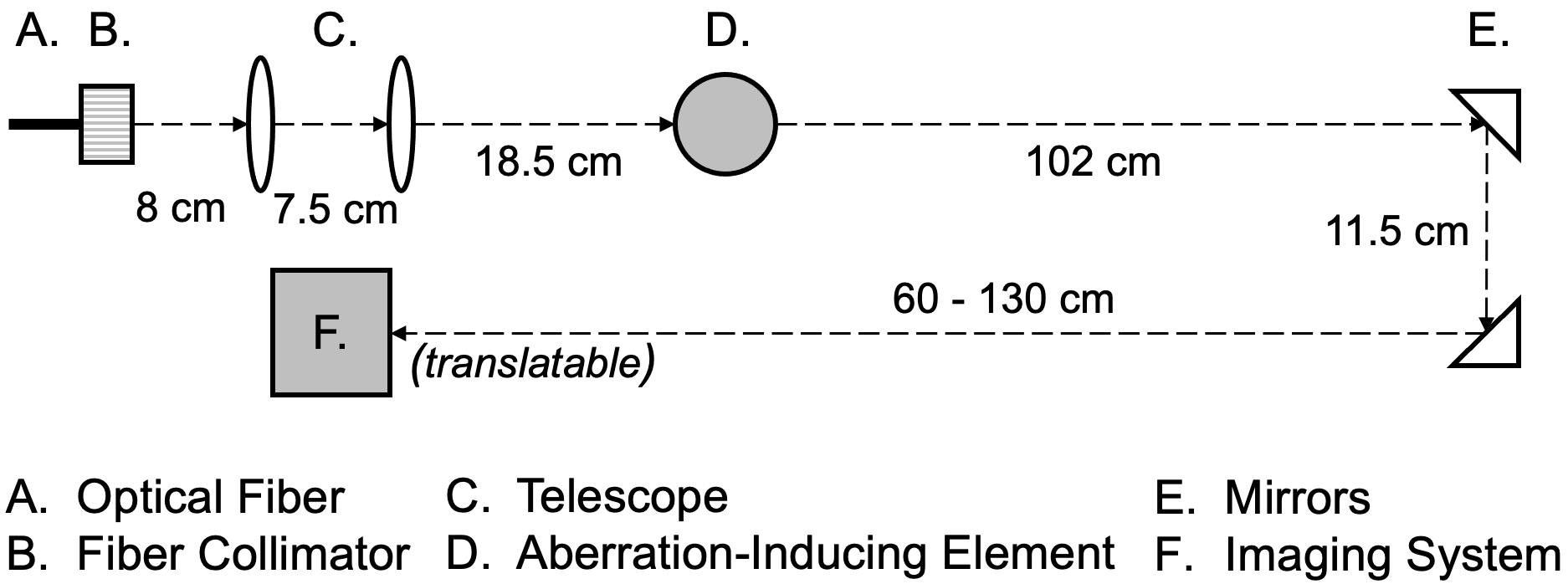}\\[0.5cm]
    \includegraphics[width = 0.4\textwidth, center]{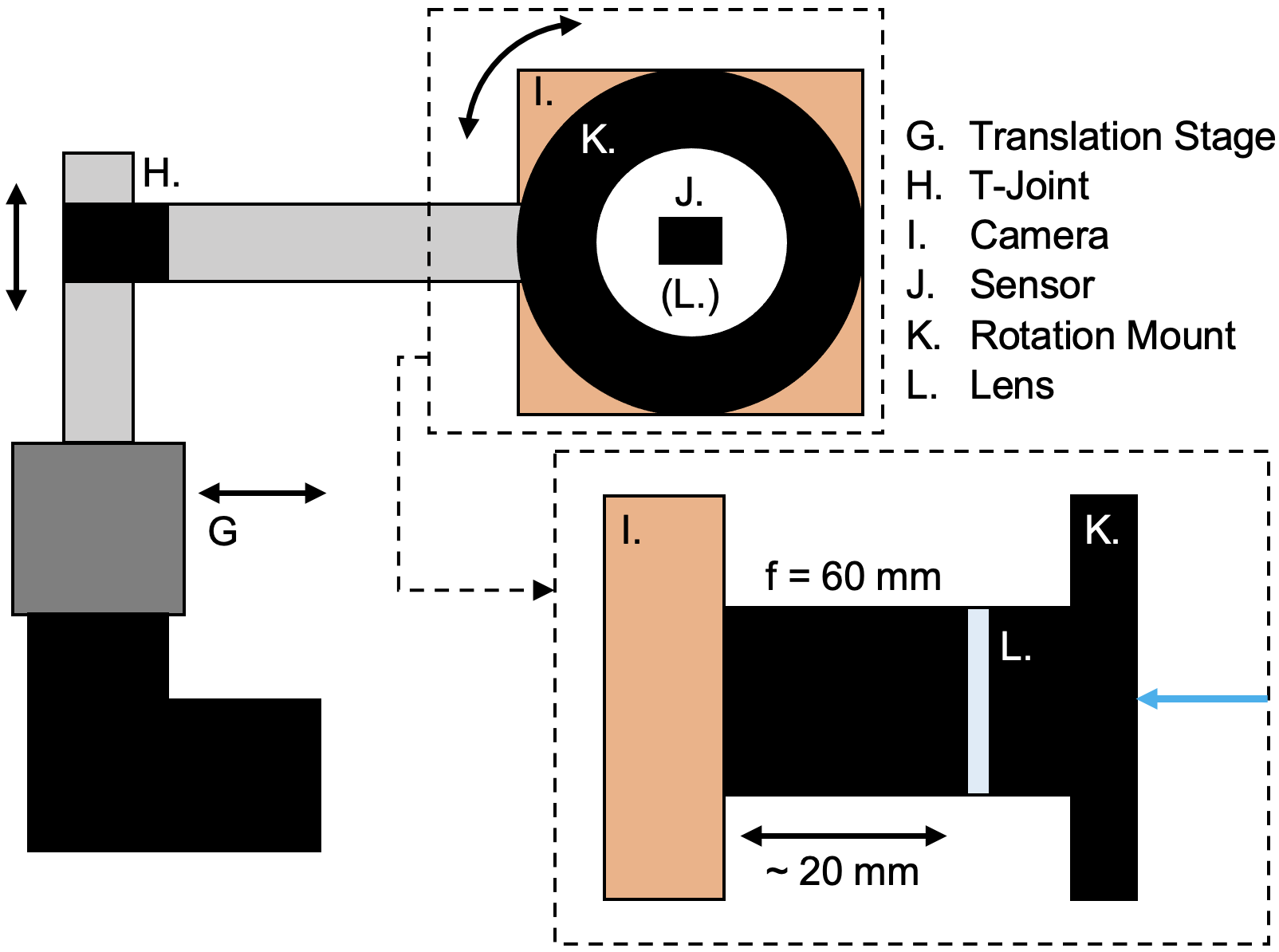}
    \caption{Experimental setup (upper) and optomechanical camera-based imaging system (lower) with two degrees of translational motion and one rotational degree of motion. In order to generate a variety of different aberrated beam profiles, the imaging system (F) in the experimental setup was moved along the beam path. A telescope with 15~mm and 75~mm lenses was used in the aperture experiment to achieve a magnification of five, while a telescope with 25.4~mm and 50~mm lenses was used in the hard-edge experiment to achieve a magnification of two.}
    \label{setup}
\end{figure}

We used an experimental setup comprised of a laser, telescope, aberration-inducing element, and a CMOS imaging system as described at the top of Fig.~\ref{setup}. We used an AOSense AOS-SP-461 diode laser to provide the 461~nm beam. We used two aberration-inducing elements for the two test cases. For the first, we used a 3.65~mm diameter Newport ID-0.5 iris diaphragm as a circular aperture and a telescope with a magnification factor of five for a beam width of 3.75~mm incident on the aperture. The iris was composed of 10 leaves, but we approximated it as circular and treated the beam profile as azimuthally symmetric. For this test case, we rotated the camera during the measurements. For the second test case, we used a Thorlabs D365 iris diaphragm to function as a hard edge by setting the iris diameter (16~mm) to be much larger than beam width (1.5~mm). We used a telescope with a magnification factor of two for these measurements. The iris was mounted on a translation stage for translation perpendicular to the beam path. The camera was translated along two axes perpendicular to the beam path for this test case. Unlike for a camera rotation during the measurement, performing the measurements with a translated camera does not assume any symmetry of the beam profile. All beam widths are given as the $1/e^2$ diameter and the iris diameters have a measurement error of $\pm$0.1~mm.

The imaging system shown on the bottom of Fig.~\ref{setup} consisted of three main components: (1) an optomechanical mount providing two degrees of translational motion in the plane perpendicular to the direction of beam propagation and one degree of rotational motion about the propagation axis, (2) a Thorlabs CS165MU1 CMOS camera with 3.726~mm by 4.968~mm image sensor and 3.45~$\mu$m pixel size, and (3) a 60~mm lens. The quantum efficiency of the Thorlabs CS165MU1 CMOS camera we used was 70\% at 461~nm. By appropriately setting the laser power, we maintained an exposure time of less than 0.5~ms. The video frame rate was 20--30~fps and was limited by the speed of the USB connection to the host computer. The raw datasets contained between 300--600 images. The 60~mm lens was mounted approximately 20~mm before the camera for slight demagnification of the beam before imaging. The effective propagation distance was changed with the addition of this lens, but this was accounted for in the simulations discussed in Section~\ref{sec:analysis}. The longest propagation distance after the aberration-inducing element that we used was 236~cm.

\subsection{Data Collection}\label{sec:data}

For the measurement, the beam profile was centered on the sensor and video capture was initiated. For the circular aperture test the camera was rotated 360$^\circ$ and for the hard-edge test the camera was translated on the order of 1~mm in each direction perpendicular to the beam path, while ensuring the beam profile remained within the image boundaries. For the circular aperture, we tested five propagation distances ranging from 177~cm to 236~cm from the aberration-inducing element by translating the camera along the beam path. The intensity of the center of the beam profile and surrounding pattern changed according to the propagation distance. We chose not to change the aperture diameter as it resulted in qualitatively redundant beam profile patterns when compared to changing the propagation distance. For the hard edge, we tested six degrees of clipping of the beam, specified by the position of the translation stage on which the hard edge was mounted. We translated the hard edge in increments of $\sim$0.4~mm across the beam profile for a total translation distance of 2~mm.

\begin{figure*}[t!]
    \centering
    \includegraphics[width = 0.9\textwidth]{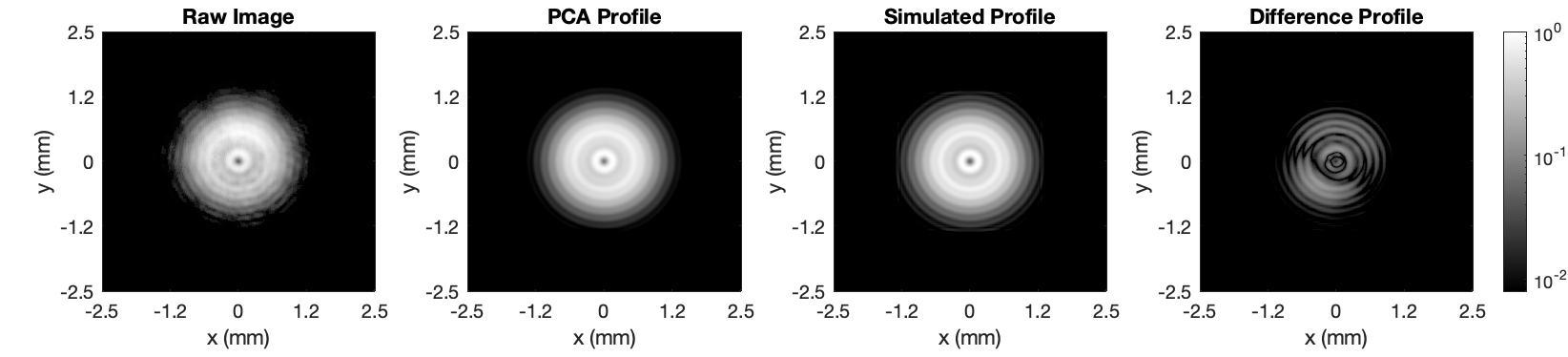}
    \includegraphics[width = 0.9\textwidth]{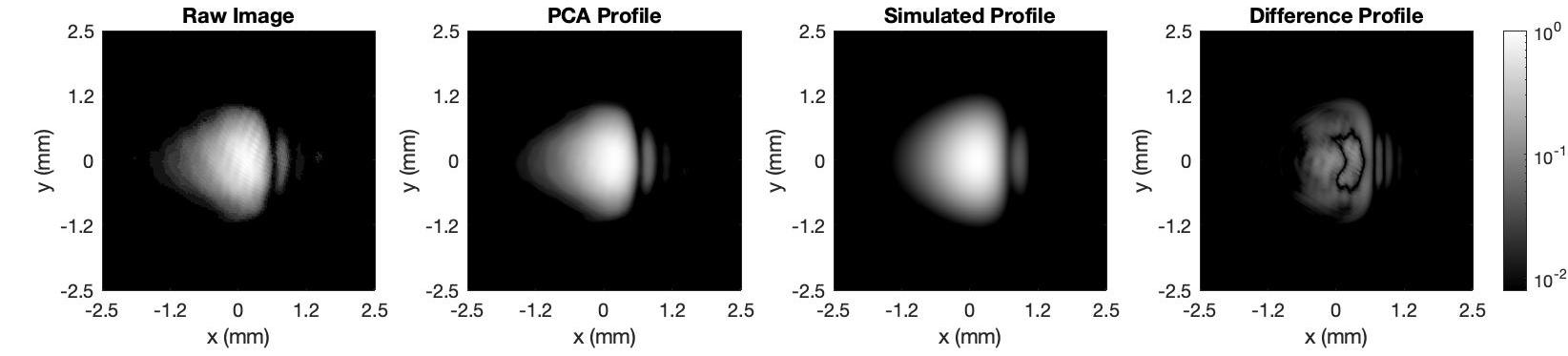}
    \caption{A sample image, normalized PCA image, simulated profile, and difference between the PCA result and simulation for two different kinds of aberrations. The first row displays the beam profile of a laser incident on a 3.65~mm circular aperture propagating 177~cm. The RMSE value of the difference image is 0.017 with a maximum pixel value difference 0.15. The second row displays the beam profile of a laser clipped on a hard-edge propagating 229~cm. The RMSE value of the difference image is 0.014 with a maximum pixel value difference of 0.13. Mean and range of RMSE values for all experiments are provided in the text. We use a logarithmic color scale here to show low-intensity details.}
    \label{diff}
\end{figure*}

\subsection{Analysis}\label{sec:analysis}

To prepare the captured images for analysis, we applied a centering algorithm consisting of region identification realized by a maximally stable extremal regions (MSER) algorithm and image translation by array padding and cropping. First, we qualitatively identified key regions in the beam profile from a sample frame of the captured video. For our tests on circular apertures, we identified the center region, either a bright or dark spot, as the key beam profile region. For our tests on hard-edges, we identified the region of highest intensity as the key region because each beam profile retained a clear maximum. 
The key regions for the rest of each dataset were then detected and extracted using a MSER algorithm. MSER algorithms are most effective at detecting ``blobs," a common characteristic of beam profiles. In our case, the MSER algorithm was executed by the MATLAB \textit{detectMSERFeatures} function \cite{Matlab, Matas2004, Nister2008}. Since the MSER algorithm generally detected many overlapping regions, we used the mean of the identified region locations. The purpose of region identification was to provide an accurate measurement of the location of the beam profile relative to the image boundaries. Misidentified frames were omitted from the final dataset by imposing a limit on the standard deviation of identified region locations. If the standard deviation of detected region locations was above a five pixel threshold, then the image was omitted from the final dataset. Typically the final datasets contained 100 frames less then the original datasets, depending on noise in the beam profile. 

For the raw images, the centering algorithm used a combination of padding and cropping of the image array in order to translate the image such that the identified center of the beam profile coincided with the center of the image. In order to ensure that we retained representative beam profiles after applying the centering algorithm, during data collection we ensured there was enough empty space around the beam profile such that the array manipulation would only affect low pixel value, near-black pixels. This requirement motivated the use of the demagnifying lens placed prior to the camera as shown in Fig.~\ref{setup}. The highest translation performed by the centering algorithm was approximately 375 pixels horizontally and 225 pixels vertically corresponding to 1.3~mm and 0.78~mm respectively.

\begin{figure*}[t!]
    \centering
    \includegraphics[width = 0.9\textwidth]{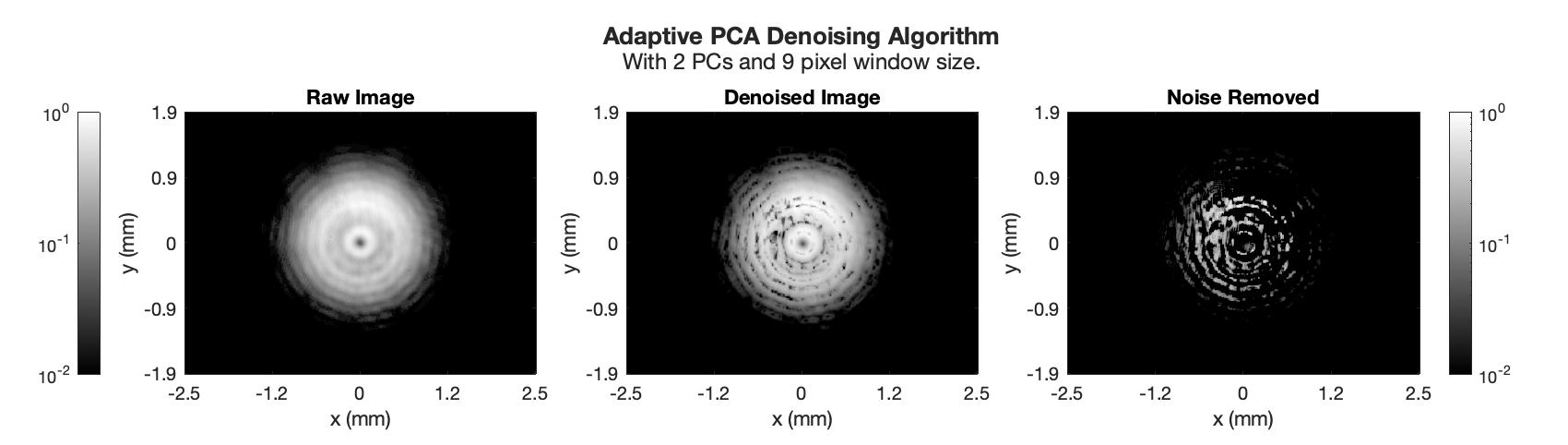}
    \includegraphics[width = 0.9\textwidth]{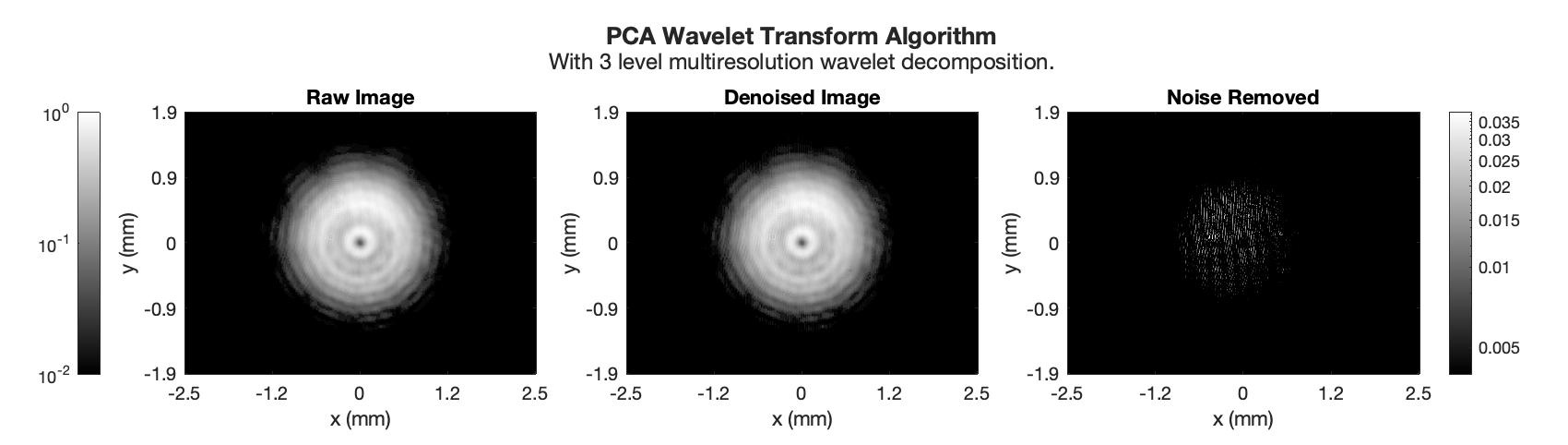}
    \caption{Application of singe-image denoising algorithms described in Ref.~\cite{Muresan2003} (first row) and Ref.~\cite{Zhang2021} (second row) as applied to the beam profile presented in the first row of Fig.~\ref{diff} (beam incident on 3.65~mm circular aperture with 177~cm of free-space propagation). In the Adaptive PCA denoising algorithm, a significant amount of noise is identified and removed; however, it misidentifies as noise a significant portion of the profile to be studied. In the PCA Wavelet Transform algorithm, the algorithm struggles to remove noise which is not purely columnar and random. This fact is reflected in the small magnitude of the noise removed. Both algorithms fail to address the diagonal fringes visible on the top right corner of the profile which are successfully removed by our method as presented in Fig.~\ref{diff}. }
    \label{alg}
\end{figure*}

After the final dataset was compiled, we performed PCA on the data. Following an approach similar to that described in Ref.~\cite{Turk1991}, we constructed a data matrix $\mathbf{X}$ from our dataset. Let $F$ be the number of images each of dimension $M \times N$. The rows of $\mathbf{X}$ corresponded to pixel values while the columns corresponded to images in the dataset, forming an $MN \times F$ array. Each image was normalized by its maximum pixel value. Next, to perform PCA we calculated the $F\times F$ covariance matrix $\boldsymbol{\sigma}$. The eigenvector of $\boldsymbol{\sigma}$ corresponding to the $k$-th largest eigenvalue is called the $k$-th principal component (PC) of $\mathbf{X}$. The $F$ eigenvectors of $\boldsymbol{\sigma}$ are mutually orthogonal; therefore, the PCs of $\mathbf{X}$ form an orthogonal basis for $F$-dimensional image space. Assuming that large variances communicate important structure, only a few PCs are needed to communicate a large amount of information about the dataset \cite{Shlens2014}. The PCA images were obtained through a linear transformation. The new pixel values were defined as a linear combination of the pixel values from the images of the original dataset, with weights prescribed by the PCs. In our case, PCA was executed by the MATLAB \textit{pca} function which uses an algorithm based on singular value decomposition.

We quantified the accuracy of the PCA result by comparing it to a simulated beam profile. The beam propagation simulation evaluates the Rayleigh-Sommerfeld diffraction integral in order to determine the beam profile after propagation. Moreover, it can incorporate various optical elements such as lenses and circular apertures. For short distances, the integral is evaluated using the angular spectrum method as described in Ref.~\cite{Goodman1996}, and for long distances it is evaluated using convolution method as described in Ref.~\cite{Shen2006}. These methods have been demonstrated to be accurate and reliable \cite{Abe2021}. The simulation generated a higher resolution beam profile than the camera, with 4,095 pixels per 4.968~mm as opposed to 1,440 for the camera. Therefore, the simulated beam profile was downsampled in order to compare it to the PCA result. The precision with which certain experimental parameters could be measured affected our ability to match the measured beam profile pattern with the simulation and fine-tuning of some simulation input parameters was required. For example, the aperture diameter was determined to about $\pm$0.1~mm, the hard-edge position $\pm$0.1~mm, and lens positions  $\pm$1~mm. Because the beam pattern was highly sensitive to these input parameters, this proved to be a limitation of the comparison.

We used the root-mean-squared error (RMSE) of the PCA-simulation difference profile as a metric to estimate the accuracy of the PCA reconstruction. The experimental results with the lowest RMSE values of the difference profiles are presented in Fig.~\ref{diff}. We analyzed the twelve total beam profiles by normalizing the PCA and simulated arrays by z-score, scaling the arrays to lie within the interval $[0,1]$, and then calculating the absolute value of the difference of the arrays. The RMSE values calculated from the normalized difference profile are presented in Table~\ref{RMSE}.


\begin{table}[htb]
    \centering
    \caption{RMSE values of the data-simulation difference profiles for a single representative raw image from each data set and for the PCA results of each data set.}
    \label{RMSE}
    \begin{tabular}{l | c c c c }
     & \multicolumn{3}{c}{RMSE} & \# of Datasets\\
     & Mean & Max & Min & \\
    \hline
    Raw Circular aperture & 0.050 & 0.088 & 0.031 & \multirow{2}{*}{5} \\
    PCA Circular aperture & 0.025 & 0.029 & 0.017 &  \\
    Raw Hard-edge & 0.028 & 0.034 & 0.024 &  \multirow{2}{*}{7} \\
    PCA Hard-edge & 0.020 & 0.035 & 0.014 &  \\
    \end{tabular}
\end{table}

The beam patterns created by the hard-edge proved to be the most sensitive to the simulation input parameters and resulted in higher RMSE values for the comparison with the PCA profile and a smaller difference between the RMSE values for the raw image and the PCA result. Nevertheless, the method did effectively remove the fringes that are apparent in the raw image.

\section{Discussion}\label{sec:discussion}

We expect that the error observed is dominated by the discrepancy between experimental and simulation parameters due to measurement error. Fine-tuning of the circular aperture diameter and hard-edge position helped to a certain degree, but parameters such as the beam envelope, precise shape of the aberration-inducing element, or lens thickness were not optimized. For example, we modeled the beam as an ideal Gaussian beam, assumed that the irises were perfect circles, and used the thin-lens approximation for the telescope and demagnifying lenses.

As a benchmark, we compared our method to the single-image denoising algorithms described in Ref.~\cite{Muresan2003} and \cite{Zhang2021}. Fig~\ref{alg} presents results as applied to one of our raw images. Ref.~\cite{Muresan2003} describes an adaptive PCA denoising algorithm in which ``training" sections of the image are sampled and analyzed via PCA before being reassembled into smaller ``denoising" sections. The algorithm best addresses zero mean Gaussian noise. Ref.~\cite{Zhang2021} describes a PCA wavelet decomposition denoising algorithm in which a three-level multiresolution wavelet decomposition is performed on the image, the dc frequency component is extracted from the vertical component of the wavelet decomposition, and then PCA is applied to extract only the denoised vertical component. The algorithm addresses columnar FPN commonly observed in images captured by CMOS image sensors. Both algorithms use PCA to target noise while retaining the important structure of the image. However, while both algorithms recognize noise present in CMOS sensors, they do not sufficiently target and remove fringes or other stationary noise present in our data.

We determine the number of PCs by comparison to the simulated beam profile. Taking RMSE values of the PCA-simulation difference profiles for 1 through 10 PCs, we found 1--3 PCs to be sufficient for the two kinds of aberrations we studied. The number of necessary PCs varies with the details of the beam profile pattern, as well as the relative intensity of noise. For various classes of aberrations, we can use results from this type of analysis to guide selection of PCs when profiling for the MAGIS-100 experiment. In general, however, the highest number of PCs should be used while sensor noise and unwanted fringes are still mitigated.

For our measurements, PCA operated similar to averaging but could also respond to non-uniformities in the dataset. All of the components of the covariance matrix were approximately equal to each other with small fluctuations caused by the movement of stationary noise with respect to the image boundary. Moreover, the sensor noise was relatively isotropic, and so for all regions in a image the centering algorithm did not alter the amount or character of sensor noise but rather shifted its detailed local structure. Given such conditions, the first PC will be a vector with components approximately equal to each other. This corresponds to PCA pixel values which are approximately an equal superposition of image pixel values.

\section{Conclusions}\label{sec:conclusions}

In MAGIS-100, beam profiling will be used primarily to (1) measure and characterize aberrations in the experimental apparatus, and to (2) test beams for a sufficiently Gaussian profile after free-space propagation. Our results suggest that the experimental beam profiling technique described is a suitable option for such applications, providing an efficient, cost-effective, and precise experimental method. We show that a CMOS image sensor combined with basic optical equipment and computer is sufficient for effective beam profiling. Furthermore, the implementation of computer vision and PCA is made simple by readily available computational packages and functions that can run easily on a laptop or tablet. Therefore, the experimental beam profiling technique described is a viable option for beam profiling efforts related to MAGIS-100 and other experiments. 

\section{Acknowledgments}

We acknowledge support from Gordon and Betty Moore Foundation Grant GBMF7945, Office of Naval Research Grant Number N00014-19-1-2181, National Institute of Standards and Technology Grant Number 60NANB19D168, and the David and Lucile Packard Foundation through a Packard Fellowship for Science and Engineering.


\bibliographystyle{ieeetr}

\end{document}